\documentclass[12pt]{article}

\usepackage{latexsym,amsmath,amssymb,pstricks,wasysym,stmaryrd,enumerate,epsfig}

\newcommand{\be}{\begin{equation}}
\newcommand{\ee}{\end{equation}}
\newcommand{\beu}{\begin{equation*}}
\newcommand{\eeu}{\end{equation*}}
\newcommand{\bea}{\begin{eqnarray}}
\newcommand{\eea}{\end{eqnarray}}
\newcommand{\beaa}{\begin{eqnarray*}}
\newcommand{\eeaa}{\end{eqnarray*}}
\newcommand{\bmx}{\begin{pmatrix}}
\newcommand{\emx}{\end{pmatrix}}

\newcommand{\del}{\partial}
\newcommand{\g}{{\frak g}}
\newcommand{\h}{{\frak h}}
\newcommand{\m}{{\frak m}}

\newcommand{\dd}{{\frak d}}
\newcommand{\tD}{{\widetilde D}}
\newcommand{\vv}{{\bf v}}
\newcommand{\vn}{{\bf n}}
\newcommand{\am}{{\alpha}}
\newcommand{\bm}{{\beta}}
\newcommand{\gm}{{\gamma}}

\newcommand{\ah}{{\hat \alpha}}
\newcommand{\bh}{{\hat \beta}}
\newcommand{\gh}{{\hat \gamma}}
\newcommand{\deh}{{\hat \delta}}
\newcommand{\eh}{{\hat \epsilon}}

\newcommand{\kh}{{\hat \kappa}}

\newcommand{\J}{{\cal J}}

\newcommand{\dxy}{\delta(x{-}y)}
\newcommand{\dpxy}{\delta'(x{-}y)}
\newcommand{\half}{\frac{1}{2}}

\newcommand{\nn}{\nonumber}

\newcommand{\eps}{\epsilon}

\advance\textheight by \topskip
\begin{document}
\baselineskip 17pt
\parindent 8pt
\parskip 9pt

\begin{flushright}
DAMTP-2004-20\\
hep-th/0402182\\[3mm]
\end{flushright}

\begin{center}
{\Large {\bf Classically integrable boundary conditions}}\\[1mm]
{\Large {\bf for symmetric-space sigma models}}\\
\vspace{0.5cm} {\large N. J. MacKay\footnote{\tt
nm15@york.ac.uk}}\\ {\em Department of Mathematics, University of
York,\\
Heslington Lane,
York YO10 5DD, UK} \\[2mm]
{\large C. A. S. Young\footnote{\tt C.A.S.Young@damtp.cam.ac.uk}}\\
{\em DAMTP, Centre for Mathematical Sciences, University of
Cambridge,\\ Wilberforce Road, Cambridge CB3 0WA, UK}
\\

\end{center}

\vskip 0.1in
 \centerline{\small\bf ABSTRACT}
\centerline{
\parbox[t]{5in}{\small
\noindent We investigate boundary conditions for the nonlinear
sigma model on the compact symmetric space $G/H$. The Poisson
brackets and the classical local conserved charges necessary for
integrability are preserved by boundary conditions which
correspond to involutions which commute with the involution
defining $H$. Applied to $SO(3)/SO(2)$, the nonlinear sigma model
on $S^2$, these yield the great circles as boundary submanifolds.
Applied to $G\times G /G$, they reproduce known results for the
principal chiral model.}}

\section{Introduction}

Over the last ten years there has been much investigation of the
boundary conditions on the half-line which preserve the
integrability of certain 1+1-dimensional field theories. However,
relatively little of this has focused on nonlinear sigma models.
There has been some work on the $O(3)$ model  (the nonlinear sigma
model on $SO(3)/SO(2)$) \cite{corr96}, and the results of
\cite{Ghoshal} on the $O(N)$ model have been extended \cite{ON}.
More recently, general boundary conditions for the principal
chiral model have been written down \cite{macka01}.

In this letter we study the classical integrability of the sigma
model on a general compact symmetric space $G/H$, and find that the
boundary conditions which preserve integrability (via the
conserved bulk charges of \cite{Evans00}) on the half-line are in
correspondence with the involutions which commute with that
defining $H$. We work in detail through the specialization of our
results to the $O(3)$ model and to the principal chiral model.

\section{The bulk model}

We first review briefly the gauged construction of the bulk $G/H$
sigma model, and its canonical structure: for the details see
\cite{Evans00,EF}. Take $G$ to be a compact Lie group, and let
$\sigma$ be an involutive automorphism of $G$ whose fixed point
set is a subgroup $H$. Then $G/H$ is symmetric and at the level of
the Lie algebras $\g=\h\oplus\m$ with
$[\h,\h]\subset\h,[\h,\m]\subset\m,[\m,\m]\subset\h$. Let $g(t,x)$
be a field taking values in $G$, and write $j_\mu=g^{-1}\del_\mu
g$. The $G/H$ model should possess the local symmetry
$g(t,x)\mapsto g(t,x)h(t,x)$: to achieve this we introduce a gauge
field $A_\mu(t,x)\in\h$ and covariant derivative $D_\mu
g\equiv\del_\mu g-gA_\mu$, so that the current \be
J_{R}^{\mu}\equiv g^{-1} D^\mu g = j^\mu-A^\mu\ee is
gauge-covariant and we can construct the gauge-invariant
Lagrangian \be {\cal L} = -\half\left< J_{R \mu}
{J_R}^\mu\right>,\label{L}\ee where $\left< \;\right>$ is a
negative-definite invariant inner product on $\g$.

The equations of motion from varying $g$ are then $D_\mu
J_R^\mu=0$, while the $A_\mu$ equations of motion impose the
constraint $J_R^\mu=0$ on $\h$. These equations of motion allow
the construction of local conserved charges whose densities are
polynomial in $J_R^\mu$ \cite{Evans00}.

The current $J_R^\mu$ is associated with the local right
$H$-symmetry. The model also has a global left $G$-symmetry, with
corresponding gauge-invariant Noether current \be J_{L \mu} \equiv
-(D_\mu g) g^{-1}.\ee We note that there exists a description of
the model purely in terms of gauge-invariant objects: setting
$q=\sigma(g)g^{-1}$, so that $q(t,x)$ is a field valued in the
`Cartan immersion'\footnote{Strictly, the Cartan immersion is
defined as $\{ g\in G\, |\, \sigma(g)=g^{-1}\}$, which may be a
finite cover of the set defined here: see the erratum to
\cite{EF}. The distinction will not be important for us in this
letter.} of $G/H$ in $G$, $\{\sigma(g)g^{-1}|g\in G\}= G/H
\hookrightarrow G$, one may re-write the Lagrangian as \be {\cal
L} =-{1\over2}\left< J_{L\mu}  J_{L}^\mu\right>
             ={1\over 8} \left< \del_\mu q^{-1} \del^\mu q \right>,\ee
since $2J^\mu_L= q^{-1} \del^\mu q $. The local charges may also
be straightforwardly constructed from $J_L^\mu$, but their Poisson
brackets \cite{forger} are then harder to handle than in the
gauged form. We therefore work with the gauged description in this
letter.

Let $\{t^a\}$ be a basis of anti-hermitian generators for $\g$,
obeying $$[t^a,t^b]=f^{abc}t^c \quad\text{and}\quad
\left<t^at^b\right>= -\delta^{ab}.$$ We may choose this basis to
be the disjoint union $\{t^\ah\}\cup\{t^\am\}$ of a basis
$\{t^\ah\}$ for $\h$ and a basis $\{t^\am\}$ for $\m$. Any
$X\in\g$ can then be decomposed as $X=X^\ah t^\ah + X^\am t^\am$.

It is convenient to eliminate the auxiliary gauge field $A_\mu$
from the Lagrangian using its equation of motion
$j^\ah_\mu=A_\mu^\ah$ before passing to phase space. The
Lagrangian is then \be \mathcal L=\half j_0^\am j_0^\am - \half
j_1^\am j_1^\am
  =\half E^{\am}_i E^{\am}_j \del_0 \phi^i \del_0 \phi^j - \half
  E^{\am}_i E^{\am}_j \del_1 \phi^i \del_1 \phi^j\ee
where $\{ \phi^i \}$ is a chart on $G$ and $E_i^a=(g^{-1}\del_i
g)^a$ are the vielbeins mapping between Lie algebra and tangent
space indices. The momentum conjugate to $\phi^i$ is \be \pi_i =
\frac{\del L}{\del (\del_0 \phi^i)} = E^{\am}_i E^{\am}_j \del_0
\phi^j,\ee and from this we define \be \J^a = E^{ai} \pi_i, \ee so
that $\J^\am = j_0^\am$, and $\J^\ah\approx 0$ must be imposed as
a constraint. One then finds the Poisson brackets \bea
    \left\{ \J^a ( x ), \J^b ( y ) \right\} & = & - f^{a b c} \J^c ( x )
    \dxy\nn\\
    \left\{ \J^a ( x ), j_1^b ( y ) \right\} & = & - f^{a b c} j_1^c ( x )
    \dxy + \delta^{a b} \dpxy \nn\\
    \left\{ j_1^a ( x ), j_1^b ( y ) \right\} & = & 0,
   \label{algebra}
\eea and the Hamiltonian density \be H\approx \half \J^\am \J^\am
+ \half j_1^\am j_1^\am = \half j_0^\am j_0^\am + \half j_1^\am
j_1^\am.\ee It follows that the constraint $\J^\ah\approx0$ is
weakly preserved under time evolution (so there are no secondary
constraints), and is first class, generating the gauge symmetry
\be \delta_{\Lambda} X(x) = - \int dy \Lambda^\ah(y)
\left\{\J^\ah(y) , X(x) \right\} \ee where $\Lambda^\ah(x)$ is
some $\h$-valued parameter which specifies the gauge
transformation. Thus the canonical formalism has been consistently
completed, and we find that the degrees of freedom of the model in
this gauged Hamiltonian description are the two $\m$-valued
currents $\J^\am=j_0^\am$ and $j_1^\am$, which transform
covariantly, together with the single $\h$-valued gauge connection
$j_1^\ah=A_1^\ah$.

The Hamiltonian above is defined only up to addition of an arbitrary
function of the constraint. We choose to set this function to be zero,
so that the Hamiltonian is well-defined, time-evolution is unique, and the
time-dependent part of the gauge freedom is fixed. The remaining
components $j_0^\ah=A_0^\ah$ are then determined by the zero-curvature
identity $\del_0 j_1 +\del_1 j_0+ [j_0,j_1] \equiv 0$:
the equations of motion following from the Hamiltonian and brackets
are compatible with this identity if and only if $j_0^\ah$ vanishes.

\section{Boundary conditions}

Consider now the model on the half-line $x\leq 0$. Demanding that the
action $S=\int_{-\infty}^0 dx \cal L$ be stationary produces the bulk
equations of motion for $x<0$, together with the boundary equation
\be\label{be}j_0^\alpha j_1^\alpha=0\ee
at $x=0$.

Let us now impose the boundary condition $g(t,0) = g_0 l(t),$ with
$g_0$ fixed and $l\in D$, where $D$ is some submanifold of $G$,
chosen (without loss of generality) such that $1\in D$. This is the
Dirichlet part of the boundary condition; the boundary equation of
motion will supplement this with a Neumann condition. Our goal is
to determine the allowed $D$.

To write the condition in terms of currents, we define \be
\dd\equiv\left\{ l^{-1} \delta l\ | l\in D, \delta l \in T_l
D\right\}\subset \g \,;\ee we shall consider only those $\dd$
which are linear subspaces of $\g$. We have $j_0\in \dd$ and,
since $j_0^\ah\equiv0$, the only non-trivial part of this
condition is \be j_0|_\m \in \dd \cap \m.\ee Let
$R=R^\top=R^{-1}:\m\rightarrow\m$ be the linear map which
restricts to $+1$ on $\mathfrak d\cap \m$ and $-1$ on $\mathfrak
d^\perp\cap\m$ -- that is, $R$ is the orthogonal reflection
through $\mathfrak d\cap \m$. The Dirichlet boundary condition
then has the form \be j_0^\am = R^{\am \bm} j_0^\bm. \ee The
boundary equation of motion (\ref{be}) then requires\be\label{cbc}
j_1^\am = - R^{\am \bm} j_1^\bm.\ee

We have said nothing yet about the conditions on $j_1^{\hat\am}$,
but these will be fixed by demanding consistency with the Poisson
brackets. First we consider the extension of the boundary
condition to the whole line by \be j_0^\am(x) = R^{\am \bm}
j_0^\bm(-x), \quad j_1^\am(x) = -R^{\am \bm}
j_1^\bm(-x).\label{ext1}\ee (This allows us to handle the
nonultralocal term in (\ref{algebra}), which is consistent with
the boundary condition since $R$ is orthogonal.) Since
$[\m,\m]\subset\h$, the brackets $\{j_0^\alpha,j_1^\beta\}$ and
$\{j_0^\alpha,j_1^\bh\}$ fix the behaviour of $j_1^\ah$: \be
j_1^\ah(x) = -S^{\ah \bh} j_1^\bh(-x)\,, \label{ext2}\ee where the
matrices $R$ and $S$ must obey \be  S^{\gh \kh} f^{\am \bm \kh} =
R^{\am \delta} R^{\bm \eps} f^{\delta \eps \gh}  \Leftrightarrow
     R^{\am \delta} R^{\bm \eps} S^{\gh \kh} f^{\delta \eps \kh} =
     f^{\am \bm \gh}.\label{cond}\ee
In a basis in which $f^{\ah \gm \delta} f^{\bh \gm \delta} =
k\delta^{\ah \bh}$, we have explicitly
\be S^{\ah \bh} \equiv k^{-1} f^{\ah \gm \delta} R^{\gm \eps}
R^{\delta \lambda} f^{\bh \eps \lambda}\label{S}\ee
(so $S$ is symmetric) and the requirement (\ref{cond}) implies
also that
\be S^2=1 \quad\text{and}\quad
         S^{\ah \deh} S^{\bh \eh} S^{\gh \kh} f^{\deh \eh \kh} =
         f^{\ah \bh \gh}.\label{cond2}\ee
Thus it emerges that the map $\tau:\g\rightarrow \g$ defined by
\be \tau: \left\{\begin{array}{cccrcl} \h & \rightarrow & \h &
\;X &\mapsto& SX \\
                                \m & \rightarrow & \m & \;X &\mapsto& RX
                                \end{array}\right. \ee
is an involutive automorphism of $\g$ which, by construction,
commutes with $\sigma$, the automorphism that defines the
symmetric target space $G/H$. On taking $\dd$ to be the
$+1$-eigenspace of $\tau$ (we had not previously specified $\dd
\cap \h$), we have that $\dd$ is a subalgebra of $\g$, $D$ is the
subgroup $\exp \dd\subset G$, and $G/D$ is itself a symmetric
space.

Now, as mentioned previously, the bulk model is known
\cite{Evans00} to possess an infinite number of local commuting
charges, of the form \be q_{\pm s} = \int dx d_{\am_1 \dots
\am_{s+1}} j_\pm^{\am_1} \dots j_\pm^{\am_{s+1}} \ee where
$d_{\am_1 \dots \am_{s+1}}$ is a symmetric tensor on $\m$
invariant under the action of the group $H$. (That is, $d_{\gm
(\am_1 \dots \am_s} f_{\am) \bh \gm} = 0$.) By arguments similar
to those in \cite{macka01}, it may be verified that, at least for
classical $G$, an infinite
subset of charges \be q_{|s|} = q_s \pm q_{-s}\ee remain conserved
and commuting in the model on the half line with boundary
conditions as above. (The crucial property needed to show this is
that for each tensor $d$, \be d_{\am_1 \dots \am_{s+1}} R^{\am_1
\bm_1} \dots R^{\am_{s+1} \bm_{s+1}} = \pm d_{\bm_1 \dots
\bm_{s+1}}.\ee For classical groups $G$, each tensor $d$ is the restriction
of a symmetric $G$-invariant tensor on $\g$ \cite{Evans00}, and so
this statement is a consequence of the more general
result \be d_{a_1 \dots a_{s+1}} \tau^{a_1 b_1} \dots
\tau^{a_{s+1} b_{s+1}}=\pm d_{b_1 \dots b_{s+1}}.\ee This is
obviously true, with positive sign, whenever $\tau$ is an inner
automorphism of $\g$, and is in fact also true when $\tau$ is outer,
for classical $G$; again, see \cite{Evans00}. We do not know how to
prove it in general.)

Our result is then that there is one classically integrable
boundary condition associated to each gauge-equivalence class of
involutive automorphisms which commute with $\sigma$.\footnote{Under gauge
transformations $g(t,x)\mapsto g(t,x) h(x)$, and in particular
$g(t,0)\mapsto g(t,0)h(0)$. It is convenient to achieve this by
requiring \be g_0 \mapsto g_0 h(0), \quad l(t)\mapsto h^{-1}(0)
l(t) h(0).\ee Then $D\mapsto h(0)^{-1} D h(0)$ and $\dd \mapsto
h(0)^{-1} \dd h(0)$. Note that the requirement that the symmetry
conditions (\ref{ext1}) and (\ref{ext2}) be preserved restricts
the allowed gauge transformations, and in particular forces
$h^{-1} \del_1 h|_{x=0}$ to vanish. Thus \emph{at the boundary}
the connection $j_1^\ah$ does transform covariantly, so there is
no inconsistency.} In gauge-invariant language, the field $q$ is
restricted to a Dirichlet submanifold
$\tD=\{\sigma(g_0l)l^{-1}g_0^{-1}| l\in D\}$ of the Cartan
immersion.

The global left $G$ symmetry $g(t,x)\mapsto g_L g(t,x)$ is broken
by this boundary condition to the remnant $g_0 D g_0^{-1}$ (which
is gauge invariant, as it should be), and with the non-local
charges forms the algebra $Y(\g,\dd)$ identified in
\cite{deliu01}.

We shall conclude with some examples, but we observe first that
two choices of boundary condition are present in all cases. If
$\tau=1$ then $D=G$ and the boundary condition is pure Neumann,
while if $\tau=\sigma$ then $D=H$ and the boundary condition is
pure Dirichlet.

\section{Example: $SO(3)/SO(2)$}

We first construct $SO(3)/SO(2)=S^2$ as the subspace $\{g\in
SO(3)|\sigma(g)=g^{-1}\}$. We choose $H=\{M_\vv(\theta)|0\leq \theta
<2\pi\}$,  where we denote by $M_\vn(\theta)$ the rotation through
angle $\theta$ about axis $\vn$, here chosen to be some fixed
$\vv$. Then $SO(3)/SO(2)=\{U\in SO(3)|MUM=U^{-1}\}$ where $
\sigma(U)=MUM$, for $M=M_\vv(\pi)$. Thus $(MU)^2=1$, so that
$MU=M_\vn(\pi)$, and we have \be\label{S2} SO(3)/SO(2)=\{
MM_\vn(\pi)| \vn.\vn=1\}.\ee

For a mixed boundary condition we choose $D$ to be any $SO(2)\subset
SO(3)$, which may be written
$$ D = \{P^{-1}M_\vv(\theta)P\,|0\leq
\theta<2\pi\},$$ and set $g_0=QP$, so that both $P$ and $Q$ are
arbitrary rotations. Then $\tau(U)= (P^{-1} M P) U (P^{-1} M P)$ and
$$ \tD=\{\sigma(g_0l)l^{-1}g_0^{-1}| l\in D\}=\{MQM_\vv(\theta)
PMP^{-1}M_\vv^{-1}(\theta)Q^{-1}|0\leq
\theta < 2\pi\}.$$ But
$QM_\vv(\theta)PMP^{-1}M_\vv^{-1}(\theta)Q^{-1}$ is conjugate to
$M$, hence squares to one, and hence is a rotation through $\pi$;
call it $M_{\vn_\theta}(\pi)$, where
$$ QM_\vv(\theta)P \quad:\quad \vv\mapsto \vn_\theta.$$
The $\vn_\theta$ form (any) $S^1\subset S^2$, but the requirement
that $\tau$ commutes with $\sigma$ fixes it to be a great circle --
that is, $P$ is a rotation through $\pi/2$.

This result should be compared with the work of Corrigan and Sheng
\cite{corr96}. They find a Lax-pair description (and hence
conserved, non-local charges) for the $SO(3)/SO(2)$ model with any
$S^1$ as the boundary Dirichlet submanifold, but explicitly leave
open the question of whether these charges are in involution. We
have found that amongst the circles only the great circles give
boundary conditions compatible with the Poisson bracket structure
inherited from the model on the whole line, and then, from the
existence of local conserved charges in involution, that these BCs
are integrable. This accords with the results of Zhao and He
\cite{zhao03}, who find that there is no consistent set of Poisson
brackets for the model with a general circle as the Dirichlet
submanifold (their `MD' condition). The case $G=SO(N)$ was also
studied in \cite{Ghoshal,ON,zhao03}.

\section{Example: the Principal Chiral Model}

The principal chiral model may be regarded as the sigma model on
$G\times G / G$, under the involution $$ \sigma \;:\quad
(g_L,g_R)\mapsto (g_R,g_L),$$ with fixed point $\{(g,g)|g\in G\}$.
Then the Cartan immersion is
$$ {G\times G\over G} = \{
\sigma(n,m)(n,m)^{-1}|n,m\in G\}=\{(mn^{-1},(mn^{-1})^{-1})|n,m\in
G\} = \{(g,g^{-1})|g\in G\}.$$

For the boundary conditions of the previous section, we require
the involutions $\tau$ which commute with $\sigma$, and write them
in terms of some non-trivial involution $\alpha$ of $G$, with
invariant subgroup $H_\alpha$. In each case we give the Dirichlet
submanifold $\tD$ in the gauge-invariant formulation, and thereby
the submanifold $\tD_G$ in the usual, ungauged formulation of the
principal chiral model, on $G$.

\noindent (i) $\tau=(1,1)$:  a pure Neumann condition, $\tD=
G\times G/G$, and $\tD_G=G$.

\noindent (ii) $\tau=\sigma$:  a pure Dirichlet condition,
$\tD=\{(e_G,e_G)\}$ and $\tD_G=\{e_G\}$, realized on the currents
$j^L_\mu=\del_\mu g g^{-1}$ and $j^R_\mu=-g^{-1}\del_\mu g$ as
$j^L_1=-j^R_1$.

 \noindent (iii) $\tau=(\alpha,\alpha)$: then
$D=H_\alpha\times H_\alpha$, and
$$ \tD=\{\sigma(g_Rl_1,g_Ll_2)(g_Rl_1,g_Ll_2)^{-1}|(l_1,l_2)\in H_\tau\} =
\{(g_Lhg_R^{-1},(g_Lhg_R^{-1})^{-1})|h=l_2l_1^{-1}\in
H_\alpha\},$$
 and so $\tD_G=g_LH_\alpha g_R^{-1}\subset G$ in the usual
formulation.

\noindent (iv) $\tau=\sigma(\alpha,\alpha)$: then $$ D=\{(s,t)\in
G\times G|(\alpha(t),\alpha(s))=(s,t)\} = \{(g, \alpha(g))|g\in
G\},$$ while $$
\tD=\{\sigma(g_Rg,g_L\alpha(g))(g_Rg,g_L\alpha(g))^{-1}|g\in G\}=
\{(g_L k g_R^{-1},(g_L k g_R^{-1})^{-1})|k=\alpha(g)g^{-1}\in
G/H_\alpha\hookrightarrow G\},
$$  so $\tD_G=g_L{G \over \; H_\alpha} g_R^{-1}\subset G$.
These results agree with those of \cite{macka01}, except that the
pure Dirichlet condition was there incorrectly identified as being
non-integrable.

\vskip 0.2in {\bf Acknowledgments}

CASY would like to thank Jonathan Evans for helpful comments on
the draft. NJM would like to thank Francis Burstall for a helpful
email discussion. CASY is funded by a PPARC research studentship.

\end{document}